\documentclass[%
 reprint,
 amsmath,amssymb,
 aps,
]{revtex4-2}

\usepackage{graphicx}
\usepackage{dcolumn}
\usepackage{float}
\usepackage{bm}
\usepackage{latexsym}
\usepackage{array}
\usepackage{amssymb}
\usepackage{amsfonts}
\usepackage{amsmath}
\usepackage{color}

\newcommand{\comment}[1]{}
\newcommand{\done}[1]{}

\newcommand{\be}{\begin{equation}}
\newcommand{\ee}{\end{equation}}
\newcommand{\bea}{\begin{eqnarray}}
\newcommand{\eea}{\end{eqnarray}}
\newcommand{\baa}{\begin{align}}
\newcommand{\eaa}{\end{align}}

\usepackage{ifthen}
\usepackage{calc}
\usepackage{url}
\usepackage{fancyvrb}
\usepackage[utf8]{inputenc}
\usepackage[english]{babel}%

\renewcommand{\selectlanguage}[1]{}

\begin{document}
\title{Computationally Predicted Electronic Properties and Energetics of Native Defects in Cubic Boron Nitride}
\author{Ngoc Linh Nguyen} \email{linh.nguyenngoc@phenikaa-uni.edu.vn}
\affiliation{Faculty of Materials Science and Engineering, PHENIKAA University, Hanoi 12116, Vietnam.}
\affiliation{PHENIKAA Research and Technology Institute (PRATI), A\&A Green Phoenix Group JSC, No. 167 Hoang Ngan, Trung Hoa, Cau Giay, Hanoi 11313, Vietnam.}
\author{Hung The Dang}
\affiliation{Faculty of Materials Science and Engineering, Phenikaa University, Hanoi 12116, Vietnam} 
\affiliation{Phenikaa Institute of Advanced Study (PIAS), Phenikaa University, Hanoi 12116, Vietnam}
\author{Tien Lam Pham}
\affiliation{Phenikaa Institute of Advanced Study (PIAS), Phenikaa University, Hanoi 12116, Vietnam}
\author{Thi Minh Hoa Nghiem}
\affiliation{Phenikaa Institute of Advanced Study (PIAS), Phenikaa University, Hanoi 12116, Vietnam}
\date{\today}

\begin{abstract}
In this study, we employ a first-principles approach to conduct a comprehensive investigation of the properties of nine common native point defects in cubic boron nitride. This analysis combines standard semi-local and dielectric hybrid density-exchange-correlation functional calculations, encompassing vacancies, interstitials, antisites, and their complexes. Our findings elucidate the influence of these defects on the structural and electronic characteristics of cubic boron nitride, such as local structures, formation energy, magnetism, and the energies of defect states within the band gap. Notably, we accurately simulate the photoluminescent spectra of cubic boron nitride induced by these defects, demonstrating excellent agreement with experimental observations. This outcome indicates that the prominent peaks in the photoluminescent spectrum at 2.5 and 2.8 eV can be attributed to the nitrogen to boron antisite (N$_{\rm B}$) and boron interstitial (B$_{\rm i}$) defects, respectively. Additionally, we investigate the energetic stability of defects under various charge states, providing valuable references for benchmarking purposes. 
\end{abstract}

\maketitle
\section{Introduction}
Cubic boron nitride (cBN) is comparable to diamond, boasting similar characteristics such as a wide band gap, exceptional temperature stability, supreme hardness, and high thermal conductivity~\cite{spriggs13PropertiesDiamond2002}. These attributes make it an enticing material that potentially replaces the costly diamond in various diamond-based technological applications~\cite{brazhkinMythsNewUltrahard2019}. However, like other semiconductors, cBN-based devices are determined by defects that are formed upon the material's synthesis. The defects significantly influence the magnetic, electronic, and optical properties of the material~\cite{nistorPointDefectsCubic2001}. Consequently, obtaining a deep understanding of the physics of different types of defects such as native defects is a crucial prerequisite for advancing optimized technologies based on cBN.

In nature, cBN and hexagonal BN (hBN) are the two most stable phase structures of BN. However, while hBN has been extensively studied in both the bulk and its exfoliated two-dimensional crystal~\cite{naclerioReviewScalableHexagonal2023}, the studies on cBN are less focused due to challenges in its synthesis conditions, such as those obtained from the conversion of hBN to cBN at high temperatures (1200--2000$^{\rm o}$C) and high pressures (2.5--7.5 GPa), and possibly requiring a catalyst~\cite{knittleExperimentalTheoreticalEquation1989, wentorfSynthesisCubicForm2004}. Recent cBN synthesis approaches have been developed, such as via chemical reactions~\cite{kaimaSynthesisCubicBoron2020} or triple direct current thermal plasma jets~\cite{ohFacileSynthesisCubic2022}, facilitating the cBN nanocrystal synthesis at lower temperatures and atmospheric pressures. These material processing methods often induce the presence of large amounts of defects, including single native, complex, and doped defects within the cBN nanocrystals created during the crystal growth, stressing the need for microscopic study for these defects. 

Experimentally, lattice defects in cBN have been studied using various methods including cathodoluminescence~\cite{tkachevCathodoluminescenceCubicBoron1985, shipiloInfluenceHighPressure1988}, X-band electron paramagnetic resonance (EPR)\cite{nistorEPRObservationFirst2000a}, near-edge x-ray absorption fine structure (NEXAFS)\cite{fosterNonempiricallyTunedRangeSeparated2012}, photoluminescence (PL) spectroscopy~\cite{berzinaPhotoluminescenceExcitationSpectroscopy2005}, and electron spin resonance (ESR)\cite{nistorPresenceDistributionImpurity2019}. These defects can be either intrinsic or intentionally created using various irradiative methods, such as neutron irradiation\cite{atobePointDefectsCubic1993}, or ion bombardment~\cite{fosterNonempiricallyTunedRangeSeparated2012}, etc. These experiments have identified the presence of different types of defects as well as their magnetic properties~\cite{clowneyGeometricParametersNucleic1996} and optical properties. However, due to the limitations in experimental resolution, it is still impossible to fully separate the experimental signals corresponding to each type of defect. In such cases, a synergy between experiments and computational methods becomes significantly important.

Theoretically, first-principles calculations for the electronic and stability of single defects have been conducted since 1980, primarily relying on low-level accurate approaches, such as Hartree-Fock~\cite{howardStabilityInterstitialsCubic1996}, tight-binding~\cite{gubanovElectronicStructureDefects1996} methods, and density functional theory (DFT) using local or semi-local density approximations (LDA) for the exchange-correlation functionals~\cite{castineiraStabilityNativeDefects1998, orellanaEnergeticsCarbonOxygen2000, gubanovElectronicStructureDefects1996, orellanaAtomicGeometryEnergetics1999, orellanaStabilityNativeDefects2001b}. Additionally, these early studies were primarily performed on small supercell models, which strongly impacted the theoretical conclusions due to finite-size effects. Highly accurate many-body perturbation theory GW methods have also been applied to study the electronic properties of cBN using the primitive cell composed of only two atoms in the basis, which is suitable only for the pristine cBN model~\cite{sattaManybodyEffectsElectronic2004}. In contrast, the extension of this approach to the defect model which often contains up to a few hundred atoms remains challenging from the computational resource. For theoretical simulating defects in semiconductors, the DFT calculation using hybrid-exchange correlation functional has been recently widely employed for different types of semiconductors. In the case of cBN, this method has been recently applied to study some potential defects serving as qubits for quantum technology applications~\cite{abtewTheoryOxygenBoronVacancy2014, bianNeutralOxygenvacancyDefect2019, turianskyTelecomwavelengthNVcenterAnalogs2023}. However, to the best of our knowledge, there is still a lack of extensive and accurate study of the native defects in cBN. This motivates us to provide a comprehensive study for the native defects in cBN using the hybrid-exchange correlation functional. The study allows us to not only predict their stability but also understand how these defects impact on electronic and optical properties of the cBN.

In this work, our study will focus on the formation energies and defect state energies of the native defects, and the defect-induced magnetism and photoluminescent spectra of cBN. The results of our study provide a deeper understanding of the behavior of native defects in cBN and offer insights into strategies for optimizing the performance of cBN-based devices. Our findings may also be relevant to the study of defects in other wide-band gap semiconductors and could pave the way for the development of novel materials with tailored defect properties for specific applications.

\section{Methodology}

\subsection{Computation setups}
The single native point defects and their complexes in the cBN are simulated by using the supercell technique where the $3 \times 3 \times 3$ supercell of the pristine cBN is constructed, containing $216$ atoms in total. The supercell imposes the defect with the concentration level of $7.8\times10^{-3}$. 
We consider 6 possible native point defects that may occur in the c-BN, including V$_{\rm N}$ and V$_{\rm B}$ vacancies, B$_{\rm N}$ and N$_{\rm B}$ antisites, and B$_{\rm i}$ and N$_{\rm i}$ interstitial defects, and 3 native defective complexes formed of the two adjacent vacancies (denoted as V$_{\rm B}$-V$_{\rm N}$), and of the two pairs of a vacancy and its adjacent antisite (denoted as V$_{\rm B}$-B$_{\rm N}$ and V$_{\rm N}$-B$_{\rm B}$). Here, the interstitial defects are formed by inserting a single B or N atom intentionally into the center of the biggest unoccupied region of the cell. 

For evaluating the stability of the defects, we consider their formation energies (E$_{\rm f}$) which is the energy cost to form a defect D in the charge state $q$, and is computed as: 
\begin{equation}{\label{formation_energy:Eq}}
    E_{\rm f}[{\rm D}^q] = E_{\rm tot}[{\rm D}^q] - E_{\rm tot}[{\rm pris}] - \sum_i \mu^{\rm c}_i N_i + q\cdot E_{\rm F} + \Delta E_{\rm corr}^q,
\end{equation}
where $E_{\rm tot}[{\rm D}^q]$ is the total energy of the supercell containing the defect D and charge $q$, and $E_{\rm tot}[{\rm pris}]$ is the total energy of the defect-free supercell. $N_i$ represents the number of atoms of type $i$ that are added (positive) or removed (negative) from the pristine system, and the chemical potential $\mu^{\rm c}_i$ represents the energy of the reservoir with which the atomic species are exchanged.
The electron chemical potential is given by the position of the Fermi level ($E_F$), taken with respect to the valence band maximum (VBM). Finally, $\Delta E_{\rm corr}^q$ is a charge-state dependent term that corrects for the finite size of the supercell, using the Freysoldt–Neugebauer–Van de Walle (FNV) correction scheme~\cite{freysoldtFullyInitioFiniteSize2009b} implemented in the CoFFEE code ~\cite{naikCoFFEECorrectionsFormation2018}.
The chemical potentials $\mu^{\rm c}_i$ are variables that represent experimental conditions. For the calculations in this
work, $\mu^{\rm c}_B$ is referenced to the total energy of a single atom
in solid-phase B, and $\mu^{\rm c}_N$ is referenced to the total energy
of an N atom in a N$_2$ molecule; we define $\Delta\mu^{\rm c}_N$ and $\Delta\mu^{\rm c}_B$
with respect to these energies. Bounds are placed on the
$\Delta\mu^{\rm c}_i$ values based on the stability condition for cBN. In
thermodynamic equilibrium:
\begin{equation}{\label{chemical_potential:Eq}}
    \Delta\mu^{\rm c}_B + \Delta\mu^{\rm c}_N = \Delta H_f ({\rm BN})
\end{equation}
where $\Delta H_f ({\rm BN})$ is the enthalpy of formation for cBN, which is predicted of 2.99 eV in good comparable with the experimental measurement of 2.77 eV~\cite{leonidovFormationEnthalpyCubic1987}.
The system in equilibrium with N$_2$ gas phase is set as an upper bound on $\Delta\mu^{\rm c}_N$ (N--rich conditions): $\Delta\mu^{\rm c}_N = 0$. Eq.~\ref{chemical_potential:Eq} then yields $\Delta\mu^{\rm c}_{\rm B} = \Delta H_f ({\rm BN})$ in the N--rich limit. Analogously, We can define
an N--poor (hereafter is called as B--rich) limit, with $\Delta\mu^{\rm c}_{\rm N} = \Delta H_f ({\rm BN})$ and $\Delta\mu^{\rm c}_{\rm B} = 0$. We note in passing that neither N--rich nor B--rich conditions realistically represent actual growth conditions, but they serve as limiting cases. 

All the calculations are done by using density functional theory (DFT) with the plane-wave and pseudo-potential methods developed in Quantum ESPRESSO package~\cite{giannozziAdvancedCapabilitiesMaterials2017, giannozziQUANTUMESPRESSOModular2009}. The structures with defects are optimized at $0$K by using the Broyden–Fletcher-Goldfarb-Shanno algorithm~\cite{fletcherNewtonLikeMethods2000}, while the size of supercell is kept fixed. The SG15 optimized norm-conserving Vanderbilt pseudo-potentials~\cite{schlipfOptimizationAlgorithmGeneration2015} are used for the electron-ion interactions. The energy cutoff for the plane wave expansion is $100$ Ry ($400$ Ry for the charge density cutoff). Brillouin-zone integration has been performed using single-$\Gamma$ point sampling and $6\times6\times6$ $k$-grid for the cBN supercell and pristine unit-cell, respectively. 
In this work, we use both the standard generalized gradient approximation with the Perdew–Burke–Ernzerhof (GGA-PBE) formalism~\cite{perdewGeneralizedGradientApproximation1996} and the dielectric-hybrid functional (DDH)~\cite{PhysRevB.93.235106} for approximating the DFT exchange and correlation functional. The later functional is adopted from the hybrid PBE0 exchange-correlation functional~\cite{doi:10.1063/1.472933} in which a portion of the exchange term is defined as the exact exchange from Hartree–Fock theory and is scaled of a constant of $\alpha = \epsilon_{\infty}^{-1}$, where $\epsilon_{\infty}$ is the macroscopic dielectric constant of the system. DDH has been shown to accurately describe the band gaps of semiconductors~\cite{PhysRevB.93.235106, doi:10.1021/acs.jctc.7b00368}, and of heterogeneous interface and surface systems~\cite{PhysRevMaterials.3.073803, PhysRevB.91.155201}. For the cBN, we adopt $\epsilon_{\infty} = 3.74$ which is the experimental macroscopic dielectric constant of the cBN bulk~\cite{Ashkenov_JAP_2003}.

\subsection{Computation verification}
We begin by verifying the accuracy of the DFT exchange-correlation functionals employed in this study. This is done by comparing the computed results of the crystal lattice constant and band gap values of the c-BN primitive cell with those measured by experiments. As presented in Table~\ref{table:structure_bandgap}, the DDH functional predicts the direct band-gap at the $\Gamma$ k-point very closely to the experimental value. It also accurately predicts the lattice constant of c-BN, with a mere 0.6\% overestimation. In contrast, the less computationally expensive GGA-PBE functional only provides reasonably accurate results for the lattice constants, with a similar order of accuracy but with underestimation. It, however, severely underestimates the band gap of c-BN, i.e. resulting in an approximately 27\% underestimation compared to experimental data. The inadequacy of the GGA-PBE functional has been extensively investigated, and it is attributed to the lack of accounting for lacking piecewise-linear condition of the total energy changing as a function of the fractional number of electrons, within this functional. In contrast, owing to the inclusion of partial exact exchange in its formulation, the DDH functional can provide quantitatively accurate estimates for the electronic properties of semiconductors. Nevertheless, due to the higher computational expense of DDH functional compared to GGA-PBE, performing structural relaxation using DDH functional becomes challenging, especially as computed on lightweight high-performing computer systems. The aforementioned results, however, motivate us to employ a combination of these two methods in our study of defect systems. In particular, all defective systems undergo thorough relaxation using the GGA-PBE functionals. Subsequently, total energy and electronic property calculations are carried out using the self-consistent field DDH functional (scf-DDH). 
\begin{table}
\begin{ruledtabular}
\caption{\label{table:structure_bandgap} Lattice constant (a) and the direct band-gap E$_{\rm g}$ value at the $\Gamma$ k-point, computed using the GGA--PBE and DDH exchange-correlation functionals. The computed values are compared with the experimental a and E$_{\rm g}$ values, respectively.}
\begin{tabular}{l l l }
          & a (\AA) & E$_{\rm g}$ (eV)\\
\hline 
 GGA--PBE& 3.62 & 4.53\\ 
 DDH& 3.59 & 6.28 \\ 
 Expt.&3.61\cite{mirkarimi_review_1997} & 6.1--6.4\cite{levinshtein_properties_2001}
\end{tabular}
\end{ruledtabular}
\end{table}

\section{Results and Discussion}
After having the established computational strategy, we carried out the calculations for structural optimization and electronic structures for the above 9 defective structures at the 5 different integer charge states, $q \in [-2, +2]$. Their formation energy and local structures of these neutral and charged defect states will be studied in detail. In particular, the optical properties such as defect state energies and orbitals relevant to the neutrally charged excitation of the optical properties, the photo-luminescence spectrum, and the charged transition energy induced by these defects will be simulated. These results provide the key aspects for controlling the performance of the cBN--based devices which are challenging to identify by using high-resolution spectroscopy experiments. 

\subsection{Energetic stability of the defects}
\begin{figure}
\includegraphics[width=0.9\columnwidth]{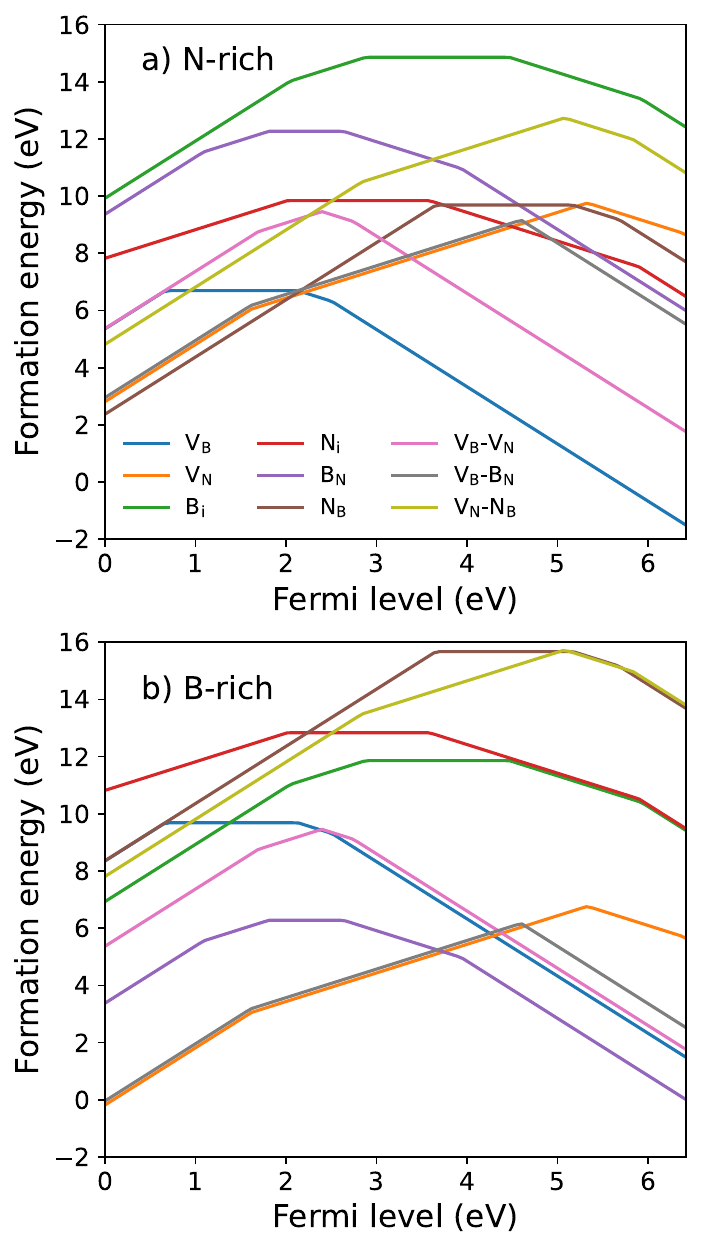}
\caption{\label{fig:defect_formation_energy} Formation energy (in eV) of the native defects in c-BN as a function of the Fermi level (in eV) under a) N-rich and b) B-rich conditions. For each defect, only the charge states with the lowest formation energies are shown.}
\end{figure}
  \begin{figure}
  \includegraphics[width=0.9\columnwidth]{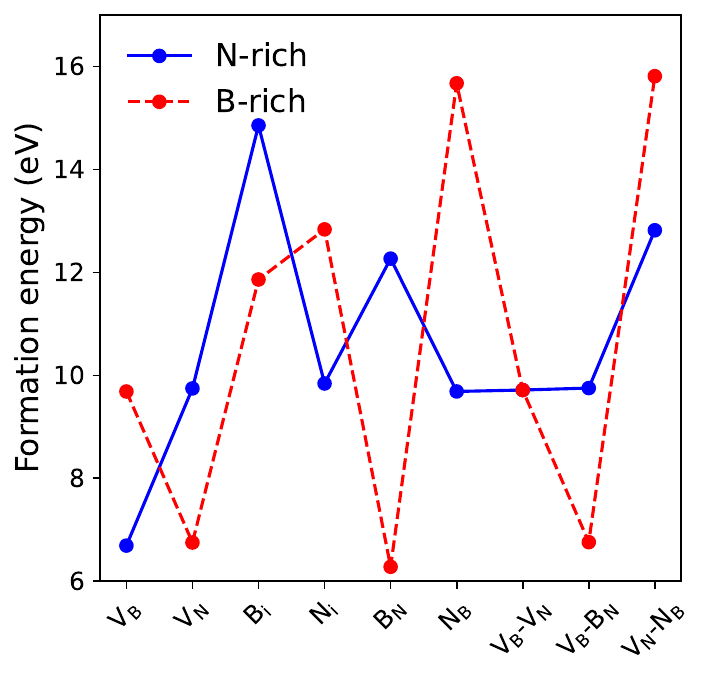}
  \caption{\label{fig:formation_energy_neutral} Formation energy (in eV) of the neutral charge native defects in cBN under N-rich (blue line) and B-rich (red dashed line) conditions.}
  \end{figure}

Defects in cBN can exist in either charged or neutral states depending on the experimental conditions. Thus, it is crucial to predict their formation energy in various charged states to accurately assess their impact on experimental observations. Previous experiments have reported the presence of some of the nine defect structures considered in this study. For instance, V$_{\rm N}$ was suspected to form through fast neutron irradiation on cBN polycrystals~\cite{Atobe_JJAP_1993} or low-energy ion bombardment on cBN powder~\cite{Peter_bombardment_2009}. Conversely, V$_{\rm B}$ formation has been observed only under high-pressure and high-temperature treatment~\cite{shipiloInfluenceHighPressure1988}. However, these measurements lack information about the defect charge states. In the subsequent discussion, we analyze the formation energy of the defects, computed according to Eq.(\ref{formation_energy:Eq}).

Firstly, we present in Fig.~\ref{fig:defect_formation_energy}-a and -b the relative $E_{\rm f}$ of the defects at different charge states as a function of $E_{\rm F}$ under the N–rich and B–rich conditions, respectively. For each defect, we present only the line segment corresponding to the charge state that gives rise to the overall lowest formation energy. 
The range value of $E_{\rm F}$ is chosen to be between zero and the theoretical cBN band gap, i.e., 6.28 eV. Under thermodynamic equilibrium, the $E_{\rm F}$ defines a probability for an electron to occupy the defect state, determining whether a defect is in either $p-$type or $n-$type. Fig. ~\ref{fig:defect_formation_energy}–a indicates that under the N--rich condition, N$_{\rm B}^{q=+1}$ has the lowest $E_{\rm f}$ and can be considered the dominant defect corresponding to a $p-$type condition as the $E_{\rm F}$ is close to the VBM. In contrast, as $E_{\rm F}$ is located from the middle of the gap to the bottom of the conduction band minimum (CBM), corresponding to an $n-$type condition, V$_{\rm B}^{q=-2}$ exhibits the lowest formation energy structure, and hence it is a dominant defect in cBN. 
On the other hand, under the B-rich conditions, Fig.~\ref{fig:defect_formation_energy}–b exhibits V$_{\rm N}^{q=+2/+1}$ and B$_{\rm N}^{q=-2/-1}$ to be the two dominant defects, corresponding to $p-$ and $n-$type, respectively. The transition energy $E_{\rm F}$ of these two states is about 3.75 eV. The above results therefore indicate that the charged carrier concentration changes from the $p-$ to $n-$types which is directly determined from the transition energy $E_{\rm F}$ changes between N-- and B--rich conditions. In comparison to the experimental findings, one can suspect that V$_{\rm N}$ and V$_{\rm B}$ under these experiments might be in the charged states, $q = +2/+1$ and $-2$, respectively, rather than being in the neutral states. These observations allow us to conclude that the conditions under which defects are charged depend on factors such as the material’s composition, structure, and environmental factors. In particular, the formation of charged defects in materials involves electron or ion concentration, doping, external conditions, chemical reactions, and radiation damage.

Secondly, we estimate $E_{\rm f}$ of the defects under neutral charge states. This estimation allows us to predict the contribution of the defects to experiments involving neutrally charged defect states, such as light absorption and photoluminescence (PL) spectroscopies. $E_{\rm f}$ of the nine defects, computed at $q=0$, are compared and presented in Fig.~\ref{fig:formation_energy_neutral}. The comparison reveals that V$_{\rm B}$ and B$_{\rm N}$ exhibit the two most energetically favorable structures under N--rich and B--rich conditions, respectively. The calculated $E_{\rm f}$ for these conditions is about 6.6 eV, closely aligning with the value of approximately 8.2 eV determined using the HSE exchange-correlation functional~\cite{turianskyProspectsNtypeConductivity2021}, thus validating the reliability of the computational method used in this work. Notably, under the N--rich condition, the formation energy of V$_{\rm B}$ is significantly lower and separated from that of the other defects, while the trio of B$_{\rm N}$, V$_{\rm N}$, and V$_{\rm B}$-B$_{\rm N}$ share closely similar formation energy. This suggests a similar likelihood of their formation under experimental conditions conducive to V$_{\rm N}$ defect discovery. Conversely, the formation energies of B$_{\rm i}$ and N$_{\rm i}$ defects exceed 10 eV under both B-- and N--rich conditions, implying their formation only under non-equilibrium conditions.

Besides, it is noteworthy that although the presence of B$_{\rm N}$ and N$_{\rm B}$ has not been experimentally identified so far, there is a common assumption that the formation of B$_{\rm N}$ and N$_{\rm B}$ in cBN might occur under excess B and N concentrations~\cite{orellanaStabilityNativeDefects2001b}. From Fig.~\ref{fig:formation_energy_neutral}, the close equivalent $E_{\rm f}$ order of B$_{\rm N}$ with that of V$_{\rm N}$ under the B--rich condition verifies this fact. In contrast, the high $E_{\rm f}$ of B$_{\rm N}$ under the N--rich condition, 12 eV, is the lowest energy in comparison to the other defects. For N$_{\rm B}$, the opposite trend is observed. Namely, $E_{\rm f}$ of N$_{\rm B}$ is as high as 9.7 and 15 eV under the N-- and B--rich conditions, respectively. 
Additionally, it is worth mentioning that the availability of native complex defects such as divacancy or vacancy-interstitial pairs has not been identified experimentally in the previous experiments. Likewise, none of the theoretical studies have reported these defects so far. However, the formation energy of these complex defects, in the order of vacancies and interstitial defects, might be possible under non-equilibrium conditions in experiments. Fig.~\ref{fig:formation_energy_neutral} also shows that except for V$_{\rm B}$-V$_{\rm N}$, the stability of the defects is strongly dependent on the experimental condition, where their formation energy and relative stability change between N-- and B--rich conditions. Here, following Eq.(\ref{chemical_potential:Eq}), removing one atom for both B and N gives the same $E_{\rm f}$ at the B--rich and N--rich conditions, respectively. Fig.~\ref{fig:formation_energy_neutral} shows that for N$_{\rm B}$ and B$_{\rm N}$, the formation energy difference between these two conditions is about 6 eV, while for other defects, it is about 3 eV. The energy changes imply that these given defects can only be found under different specific conditions.

\subsection{Structural properties of the neutral defects}

Next, we discuss the structural properties of the considered defects. In principle, the inevitable distortion of the pristine lattices induced by the presence of defects has a significant influence on the mechanical and electronic properties of materials~\cite{nguyen_understanding_2023, nguyen_understanding_2023_1}. However, experimental measurements of atomic structure and bond lengths directly obtained around a native defect are challenging. Only in the case of impurities with relatively heavy atomic mass can such measurements be obtained directly from extended X-ray absorption fine structure~\cite{lee_extended_1981, freysoldt_first-principles_2014}. This challenge necessitates a synergy between experiment and atomistic modeling to reveal local structure information around native defects. From our calculations, we observe that the nine native defects in cBN induce a strong distortion in the crystal symmetry of the defect’s nearest neighboring atoms. However, although both B-- and N--atoms have the same symmetry, with each atom having four nearest neighbors, due to the discrepancy in their atomic radii, one might expect that the possibility of forming B--related and N--related single defects, as well as the impact on their local structures, differs. Additionally, the defects' charge states also impact their structures differently, where charging only affects locally in V$_{\rm B}$ and V$_{\rm N}$ structures, while for the others, it is almost unchanged with respect to the charge numbers. In the following, we examine in detail these discrepancies for each type of defect.

Regarding structural impacts induced by the two vacancy defects, V$_{\rm N}$ and V$_{\rm B}$, our calculations indicate that V$_{\rm N}$ can affect its local structures more strongly than V$_{\rm B}$. V$_{\rm N}$ draws the nearest neighbor B atoms closer to its center, resulting in an average distance between the vacancy center and the 4--nearest B neighbors about 10\% shorter than the average B--N bond length. Conversely, the presence of a V$_{\rm B}$ center induces only a 6\% contraction. This phenomenon reflects the strong hybridization between unpaired electrons on the N atoms upon removal of the V$_{\rm B}$. On the other hand, electronic charging also impacts these defects differently. Namely, while positive charging to V$_{\rm B}$ tends to increase the local distortion, it restores the local distortion of V$_{\rm B}$ to the pristine one upon negative charging. In contrast, the opposite behavior is found in V$_{\rm N}$. The interpretation of this impact of the charging to the structure is reflected via the occupancy of the electron on the localized defect orbitals, which is discussed in Section~\ref{sec:section2}.

Regarding B$_{\rm i}$ and N$_{\rm i}$ defects, our calculations reveal that the lowest energy configurations correspond to the interstitial B and N atoms allocated at the largest unoccupied region of the supercell. This conclusion is reached by comparing the total energy between B$_{\rm i}$ and N$_{\rm i}$ optimized from different possible initial positions of the interstitial B and N atoms in the cell. We find that in the B$_{\rm i}$ structure, the formation of a B--B pair occurs between the B interstitial and its closest B site in the lattice, with the B--B bond length approximately 0.5~\AA, while in the N$_{\rm i}$ structure, the N atom of the pair binds with its two first--neighbor N atoms with equal bond length, with the N--N bond length approximately 0.7~\AA.

For B$_{\rm N}$ and N$_{\rm B}$, their effects on the local structures of the cBN crystal are different. The equilibrium geometry of B$_{\rm N}$ shows an outward breathing relaxation of the neighboring B atoms, with a small distortion of the B antisite along the [100] direction approximately 0.82~\AA~from the N site. The distance between B--N and its first neighbors is 5.2\% larger than in the unrelaxed system, while the distances between first and second neighbors are 0.2\% shorter. No distortion along the [111] direction has been detected. For N$_{\rm B}$, we observe an off-center distortion of the N antisite along the [100] direction, with a minimum at 0.35~\AA~from the B site. The N antisite binds to two neighboring N atoms, forming a bridge structure with equal N--N bond lengths of 1.41~\AA, and an angle of 122${\rm ^o}$ between the bonds. We find that these impacts are less pronounced than those studied in the previous work~\cite{orellanaAtomicGeometryEnergetics1999}, in which the same GGA-PBE exchange-correlation functional was used, yet a smaller supercell size was considered. This discrepancy highlights the need for a supercell size convergence check for such structural examination.

For the trio of complex defects, their impacts on the localized structure can extend to a larger size compared to the single defect. In the case of the divacancy V$_{\rm B}$-V$_{\rm N}$, it creates a hollow in the crystal, with a volume formed by its nearest neighboring atoms smaller than the original one. Here, the nearest neighboring atoms are drawn toward the vacancy centers. We find that such bonding reductions are only 8.5\% and 5.4\% less pronounced than those of the single V$_{\rm B}$ and V$_{\rm N}$, respectively, implying a compensating effect between these two defects in the case of the V$_{\rm B}$-V$_{\rm N}$ complex. Regarding the cases of V$_{\rm B}$-B$_{\rm N}$ and V$_{\rm B}$-N$_{\rm B}$, we find that the former defect can induce a contraction between these two types of defects: while the V$_{\rm B}$ causes its neighboring atoms to be drawn towards the center, the antisites B$_{\rm N}$ and N$_{\rm B}$ tend to push their neighboring atoms further from the defect centers. These two opposite effects lead to relative changes in bond length (5.7--6.8\%) and bond angle (7.8--6.8\%) with respect to those of the pristine cBN crystal. Here, the V$_{\rm B}$-N$_{\rm B}$ complex has a stronger impact on the structure than that of V$_{\rm B}$-B$_{\rm N}$, reflecting the fact that the radius of the N atom is larger than that of the B atom. Detailed atomic coordinates of all GGA-PBE optimized structures are presented in Supplementary Material (SM).~\cite{Supplemental_material}.

\subsection{Electronic properties and formation of the defect states induced by the neutral defects} \label{sec:section2}
\begin{table}
\begin{ruledtabular}
\caption{\label{tabe:spin_state} Ground-state spin of the c-BN defective structures computed at different charged states. Only the spin states localized on the defect centers are presented.}
\begin{tabular}{l r r r r r}
          & +2 & +1 & 0 & -1 & -2 \\
\hline 
 V$_{\rm B}$& - & 1 & 3/2 & 1& 1/2\\ 
 V$_{\rm N}$& 1/2 & 0 & 1/2 & 1& -\\
 B$_{\rm i}$& 1/2 & 0 & 1/2 & 0& 0\\ 
 N$_{\rm i}$& -   & 1 & 1/2 & 0& 0\\ 
 B$_{\rm N}$& -   & 3/2 & 1 & -& 0\\ 
 N$_{\rm B}$& 0   & 1/2 & 0 & 1/2& -\\ 
 V$_{\rm B}$--V$_{\rm N}$& - & 1/2 & 1 & 1/2& 0\\ 
 V$_{\rm B}$--B$_{\rm N}$& 1/2& 0 & 1/2 & 1& -\\ 
 V$_{\rm N}$--N$_{\rm B}$& 1/2& 0 & 1/2 & 0& -\\ 
\end{tabular}
\end{ruledtabular}
\end{table}
  \begin{figure*}
  \includegraphics[width=1.9\columnwidth]{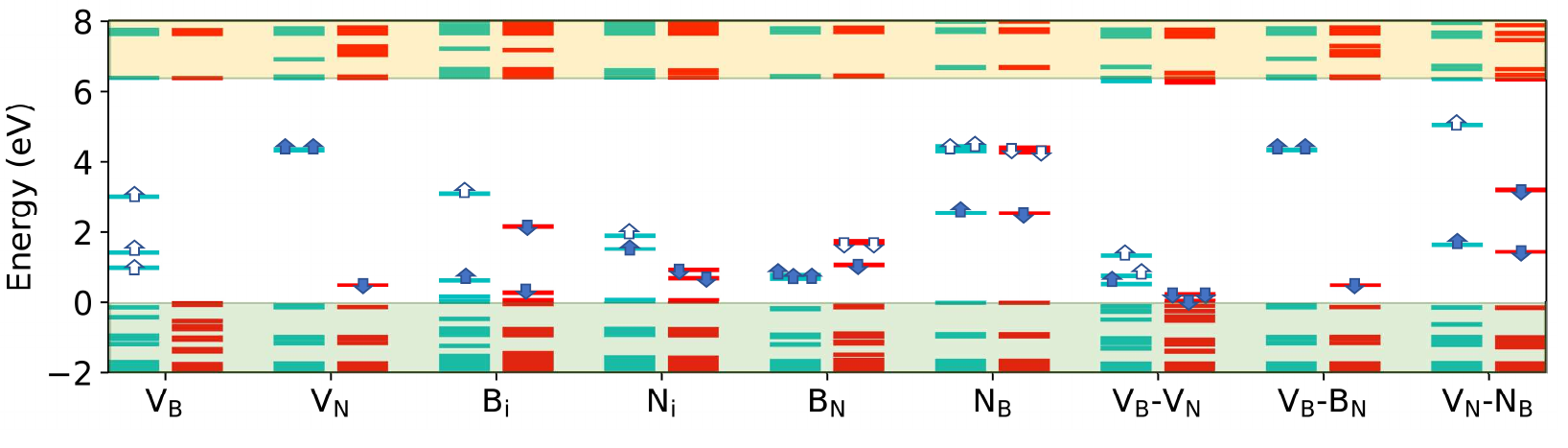}
  \caption{\label{fig:defect_state_energy} The electronic structures of the nine c-BN defect models. The deep defect states are denoted with black or white arrows corresponding to occupied or unoccupied states, respectively. The up (down) arrow corresponds to the spin channel up (down) of the Kohn-Sham states, respectively. The valence-band and conduction-band energy regimes are highlighted in blue and orange regimes respectively.}
  \end{figure*}

The formation of defect states within the band gap of an intrinsic semiconductor profoundly influences its electronic and optical properties. Furthermore, their energy position relative to the CBM and VBM, along with their electron occupancy, characterizes the conductivity of the material, determining whether it becomes $n-$type, $p-$type, or bipolar semiconductor. 
In this section, we delve into how native defects modify the electronic properties of cBN. Our calculations encompass both neutral and charged systems. Fig.~\ref{fig:defect_state_energy} illustrates the energy of defect states relative to the VBM and their corresponding electron occupancy for the neutral system. It becomes evident that V$_{\rm B}$ and V$_{\rm B}$- B$_{\rm N}$ transform cBN into n-type and p-type semiconductors, respectively, while other defects lead to the formation of occupied and unoccupied defect states within the band gaps, characterizing a bipolar semiconductor.
Moreover, the presence of defects induces a transition in cBN from non-spin polarization to spin polarization, thereby magnetizing the system. As demonstrated in Table~\ref{tabe:spin_state}, the multiple spin values of cBN across the different defects and their corresponding charge states highlight their potential applications in various fields such as dilute magnetic semiconductors and spintronic devices. Furthermore, it is worth stressing that the finite spin value ($>0$) of the defects states and their possible stability, implying that these defects can be considered as potential quantum bits used within the quantum technologies besides of the intentionally doped complexes such as V$_{\rm B}$-O$_{\rm N}$~\cite{abtewTheoryOxygenBoronVacancy2014, bianNeutralOxygenvacancyDefect2019},  V$_{\rm B}$-C$_{\rm B}$~\cite{turianskyTelecomwavelengthNVcenterAnalogs2023}, and V$_{\rm B}$-Si$_{\rm B}$~\cite{turianskyTelecomwavelengthNVcenterAnalogs2023}. In the following, we provide a detailed discussion on the electronic structures of the neutral defects in the groups to reveal the role of each defect.

{\it Boron and Nitrogen vacancies:}  
Fig.~\ref{fig:defect_state_energy} shows that both V$_{\rm B}$ and V$_{\rm N}$ induce the formation of the defect states within the cBN band-gap. For V$_{\rm B}$ defect, it forms the three spin-polarized defect states in the majority spin channel. They are unoccupied (i.e. $p$--type), degenerated in energy, and characterized as $p$--like t$_2$ states originating from N atomic orbitals. These states are positioned approximately 1.1, 1.3, and 2.5 eV above the VBM. The strong energy splitting between these three states accompanies a slight breaking of symmetry in the defect orbitals due to the non-locality of the exchange-potential operator within the DDH formalism~\cite{engelExactExchangeRelativistic2008}. Interestingly, these orbitals align through $p-p$ orbital coupling mediated by hole carriers at the defect sites, resulting in magnetic moments aligned along the [111] direction. The spin polarization gives rise to ferromagnetic behavior in the system, with a total magnetization and absolute magnetization of 3$\mu{\rm_B}$, where $\mu{\rm_B}$ is the Bohr magneton constant. For V$_{\rm N}$ defect, it induces the formation of three distinct $n$-type defect states within the cBN band gap: two of them are on the major-spin channel and situated about 1.7 eV below the conduction band minimum (CBM), one is on the minority spin channel with the energy of 0.4 eV above the VBM. One can see that the former two states play the role of shallow donor states for cBN. The isodensity analysis for these defect states exhibits that the orbitals of the former states resemble the e-like orbital, while the latter resembles the $s-s$-like a$_1$ orbital. The asymmetric occupation distribution between the two spin channels induces the localized spin-polarization with a spin magnetization of 1$\mu{\rm_B}$.

Both the above properties of V$_{\rm B}$ and V$_{\rm N}$ defects were previously predicted by using the GGA-PBE functional on a smaller supercell (64 atoms)~\cite{gubanovElectronicStructureDefects1996}. These computational factors lead to different conclusions obtained in this work as compared to those in the previous one. For instance, with GGA-PBE, the band gap is underestimated, and the shallow donor states are predicted to be above the CBM, resulting in an overestimation of the Fermi level. The impact of supercell size is evident in the position of the three occupied defect states relative to the unoccupied ones. In a smaller supercell, these states have energy levels in the same position as the unoccupied ones and are situated around 1 eV above the VBM. However, in larger supercells, the unoccupied states shift downward, becoming closer to the VBM. This size-dependent effect is also observed in other (III) nitride semiconductors~\cite{vandewalleFirstprinciplesCalculationsDefects2004}

{\it Boron and Nitrogen interstitials:} Fig.~\ref{fig:defect_state_energy} indicates that each B$_{\rm i}$ and N$_{\rm i}$ defect forms four spin-singlet levels within the band gap, where the two lower energy states are fully occupied and they are spin broken symmetry; the two higher energy are spin-polarized states where one of them are fully occupied and the other is empty. One can find that the spin-broken symmetry is more pronounced in the case of B$_{\rm i}$, which is explained as the more asymmetric of the B$_{\rm i}$ within the crystals. Here, the spin-broken symmetry in both systems exhibits the localized spin-polarization with a spin magnetization of 1$\mu{\rm_B}$. 

{\it Boron and Nitrogen antisites:} Regarding impacts on electronic structure induced by these two vacancy defects, Fig.~\ref{fig:defect_state_energy} shows that B$_{\rm N}$ induces the formation of the 6 spin-polarization defect energy states within the cBN band gap. Here, three of them are fully occupied and stay in the spin-major channel and they have close energy about 1.2 eV above the VBM. In contrast, on the spin minority channel, they are composed of two unoccupied states with an energy of about 2 eV above VBM, and one occupied with an energy of about 1.3 eV above VBM. The iso-density plots of these defect states indicate that the two 4 occupied orbitals are characterized by the $p$--like $t_2$ state, while the two unoccupied ones are characterized by the excited $e$ state. We find that this spin-broken symmetry goes along with the larger Jahn-Teller distortion~\cite{koppelJahnTellerEffectFundamentals2009} compared to the GGA-PBE results. Also, it induces the spin polarization with the spin magnetization of 2$\mu{\rm_B}$ in B${\rm _N}$. 
In contrast, the N${\rm _B}$ defect does not yield spin polarization in the system. This defect induces the formation of the six defect states in the cBN band gap, where each spin channel has one fully occupied state and two unoccupied states. From Fig.~\ref{fig:defect_state_energy}, one can find that the two occupied states are in the mid-gap, i.e. about 3.7 eV above the VBM, while the unoccupied states are slightly higher, i.e. about 4 eV above the VBM. By analyzing the orbital iso-density of these states, we find that the occupied state is characterized by $s-$like $a_1$ state in the upper part of the band gap, while the two unoccupied states characterized by the $a_1$-state originated from the split of a $t_2$ state. 

In comparison with the previous theoretical work~\cite{orellanaStabilityNativeDefects2001b} done using the GGA-PBE functional for B$_{\rm N}$ and N$_{\rm B}$, we find that while the orbital characteristics of the defect states are the same, while their energy distribution results are significantly different. This result again highlights the need to have a high-level approximation for the exchange-correlation functional for predicting the defect state energies and band gap of cBN. 

{\it The vacancy and interstitial complex defects:} From Fig.~\ref{fig:defect_state_energy}, one can observe that the distribution of defect state energies differs with respect to the different defects. For V$_{\rm B}$--V$_{\rm N}$, it gives rise to the 6 defect states within the cBN band gap. 
On the spin-majority channel, there are three states composed of one fully occupied state and two degenerate unoccupied states with energies of 0.7, 0.8, and 1.2 eV above the VBM, respectively. 
On the spin-minority channel, all three states are fully occupied and have energies close to the VBM, i.e., $<0.2$ eV. 
By analyzing the isodensity of these defective orbitals, we find that they are still localized on the defect center, resembling the $p$-like characters of the single V$_{\rm B}$ or V$_{\rm N}$, yet the level of localization is less than that of the defect orbitals of the single V$_{\rm B}$ or V$_{\rm N}$ due to interaction between them and the host orbitals. 
Such interaction is reflected in the close energy levels between them and the VBM. The spin-broken symmetry induces spin polarization, resulting in the system's paramagnetic properties with a magnetization of 2$\mu_B$.
For V$_{\rm B}$--B$_{\rm B}$ and V$_{\rm N}$--N$_{\rm B}$, we find that although they yield the same spin magnetization of 2$\mu_B$, their defect energy and electron occupancy on these states are different.
Fig.~\ref{fig:defect_state_energy} shows that V$_{\rm B}$--B$_{\rm B}$ is characterized by the $n$-type defects with three fully occupied defect states within the band gap, where the two states on the spin-majority channel have energies about 2 eV below the CBM, and the one on the spin-minority channel has energy about 0.2 eV above the VBM.
On the other hand, for V$_{\rm N}$--N$_{\rm B}$ defect, it creates four defect states within the band gap, which is composed of two states on the spin-majority channel (one occupied and one unoccupied state), and two occupied states on the spin-minority channel. 

Furthermore, it is worth stressing that due to the localization of the defect states, the charging or discharging of the systems corresponds to adding/removing consequently electrons to/from these states, respectively. 

\subsection{Simulation of cBN photoluminescence spectra}
The defect states within the band gap of semiconductors serve as optical centers. In the presence of external excitation sources such as thermal energy, lasers, or ultraviolet light, electrons from the valence band or donor states are excited to the conduction band, leaving behind a hole. The excited electron then undergoes non-radiative decay to reach the conduction band minimum (CBM) before recombining with the hole, simultaneously emitting a photon. This recombination mechanism can occur between band and defect states or between band and band states. In this work, we employ a model based on band and defect states photoluminescence (PL) spectra to simulate the PL spectra of cBN induced by the nine defects considered in this study. The theoretical PL spectra are simulated following Fermi’s golden rule~\cite{Bebb1972}, which describes the radiative recombination between an excited electron located at the CBM, denoted as $\varphi_{\rm CBM}$, and a hole residing at the deep defect state, denoted as $\varphi_{d}$, with an energy level within the band gap of the semiconductor:
\begin{equation}{\label{eq:PL}}
    PL(\omega) = \sum_d \mathcal{N} \omega^2 |\langle \varphi_{\rm CBM} |\hat{r}| \varphi_{\rm d}\rangle|^2 \times \delta(\epsilon_{\rm CBM} - \epsilon_{\rm d} - \omega)
\end{equation}
Here, $\hat{r}$ is the position operator, $\epsilon_{\rm CBM}$ and $\epsilon_{\rm d}$ are the energy levels at the CBM and defect state, respectively. $\mathcal{N}$ denotes the normalization factor.

In our calculations, the single-particle states $\varphi$'s and the energy levels $\epsilon$'s in Eq.\eqref{eq:PL} are obtained from the solutions of the Kohn-Sham equation in the DFT calculations. We note that the complete formula for the PL spectra \cite{Bebb1972} includes the electron (hole) occupancy in the initial (final) state. However, since there are only deep defect states, the excited electrons remain predominantly around the CBM, while the hole occupancy at the defect states is always unity, justifying Eqn.\eqref{eq:PL}.

\begin{figure}
  \includegraphics[width=0.9\columnwidth]{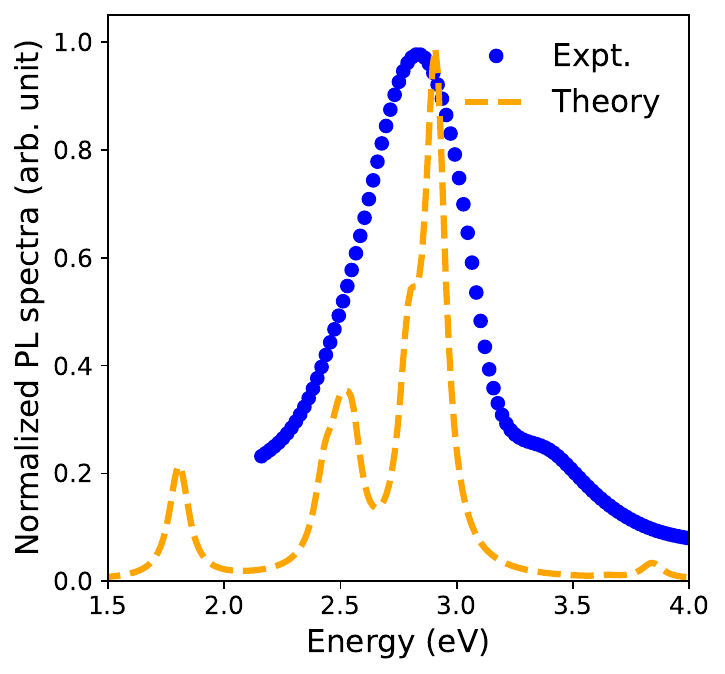}
  \caption{\label{fig:pl_spectra} The total photoluminescence spectra computed as averaging over the spectra computed for the single-defect geometries (orange dashed line) and compared with the experimental spectrum (blue dots) taken from Ref.~\onlinecite{berzina_photoluminescence_2005}.}
  \end{figure}

The PL spectrum for each defect is computed following Eq.\eqref{eq:PL} and presented in the SM~\cite{Supplemental_material}. In Fig.~\ref{fig:pl_spectra}, we display the averaged PL spectra over the 9 defects, with the averaging weights determined by the Boltzmann distribution. The obtained result demonstrates that the theoretical spectrum aligns very well with the experimental one. Our calculation suggests that the main peaks in the experimental spectra may be attributed to the PL originating from the two defects N$_{\rm B}$ and V$_{\rm B}$. Interestingly, we also observe a small peak around 1.7 eV corresponding to the V$_{\rm N}$ defect. This energy falls in the far-red range and could be advantageous for producing white-light-emitting diodes. It's worth noting that the highest peak in the experimental spectrum is broader than in the theoretical ones. This could be rectified by exploring other types of defects. A more comprehensive study of PL spectra for different defect types will be considered in future steps.

\subsection{Charge-state transition level of the defects}

\begin{figure}
\includegraphics[width=0.9\columnwidth]{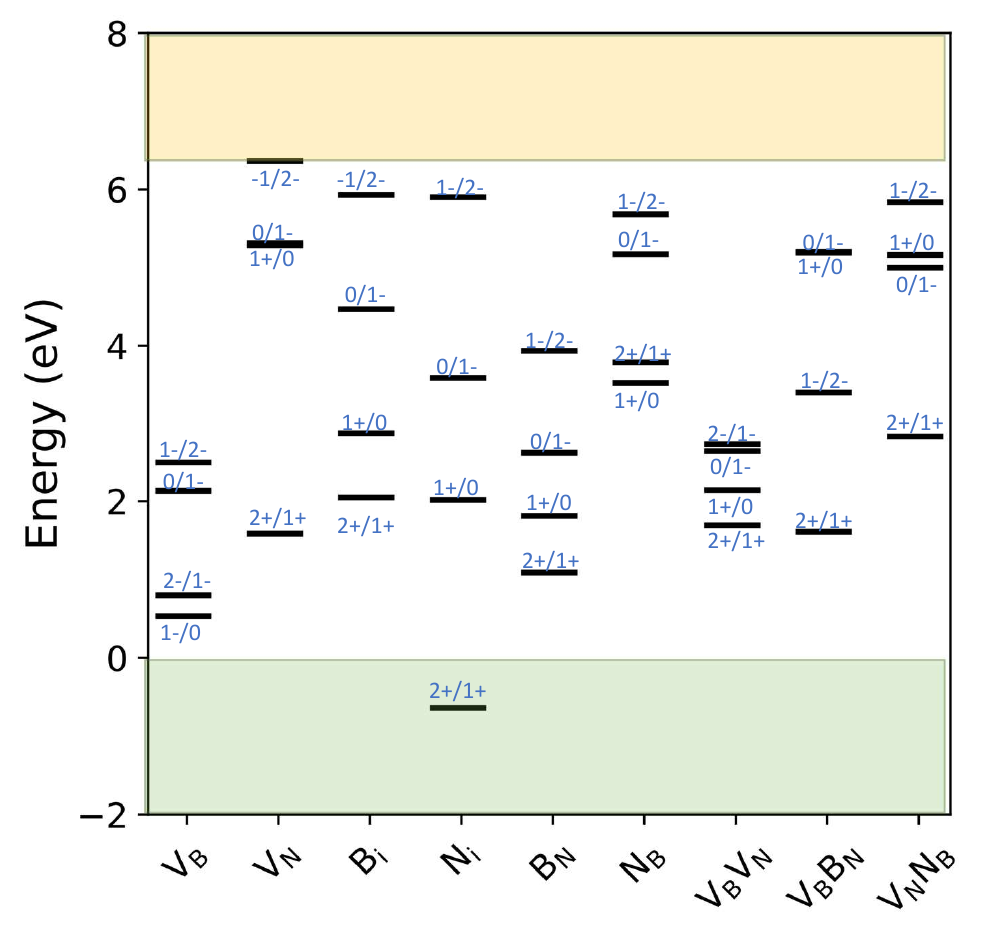}
\caption{\label{fig:defect_transition_levels} Charge-state transition levels for the native defects in c-BN. The valence-band and conduction-band energy regimes are highlighted in blue and orange regimes respectively.}
\end{figure}

Lastly, we discuss about the charge-state transition level $\epsilon(q|{q'})$ of the defect. Also, this quantity allows us to determine the stability of the charged defects. Theoretically, this quantity is defined as the Fermi-level position below which the defect is stable in the charge state $q$, and above which it is stable in charge state ${q'}$, and is calculated as:
\begin{equation}{\label{transition_level:Eq}}
    \epsilon(q|{q'}) = \frac{E_{\rm f}[{\rm D}^{q'}, E_{\rm F} = 0] - E_{\rm f}[{\rm D}^{q}, E_{\rm F} = 0] }{{q'} - q}
\end{equation}
where $E_{\rm f}[{\rm D}^{q'}, E_{\rm F} = 0]$ is the formation energy of ${\rm D}^{q'}$ when the Fermi level is at the VBM (i.e, for $E_{\rm F} = 0$). Experimentally, $\epsilon(q|{q'})$ can be obtained in the deep-level transition spectroscopy experiments, or they can be derived from an analysis of temperature-dependent Hall effect data~\cite{vandewalleFirstprinciplesCalculationsDefects2004} in the case of shallow center with respect to the crystal surfaces.  

The charge-state transition levels of each defect have been computed and are illustrated in Fig.~\ref{fig:defect_transition_levels}. It is observed that all $\epsilon(q|{q'})$ values between the successive $q$ and ${q'}$ states fall within the range of the VBM and CBM energies, except for the cases of $\epsilon(+2|{+1})$ for N${\rm i}$, which lies approximately 0.5 eV below the VBM, and $\epsilon(-1|{+2})$ for V${\rm N}$, which is about 0.01 eV above the CBM. This distribution of $\epsilon(q|{q'})$ signifies the phase transitions occurring between the most stable charged defect structures as the Fermi energy ($E_{\rm F}$) varies from 0 to the energy band gap (E$_{\rm g}$). Given the absence of prior experimental and theoretical reports on this parameter, the provision of these measurements is both valuable and essential for guiding similar studies.

\section{Conclusion}
In summary, our study involved conducting extensive simulations to investigate the electronic structures and energetic stability of the nine native defects within cBN. We employed first-principles DFT calculations, utilizing the combination of the less computationally expensive standard semi-local and highly accurate DDH exchange-correlation functionals. We meticulously assessed the accuracy of these functionals, confirming their ability not only to predict structural properties but also the band gap of the cBN systems, aligning well with experimental observations. Notably, the use of the Kohn-Sham CBM and defects states computed with the DDH functional enables us to precisely simulate the PL spectrum of cBN, through the Fermi golden rule formalism. The agreement between the theoretical and experimental spectra offers a precise interpretation of the origins of the excitation energy observed in experiments and provides confident predictions for the excitation energy in previously unexplored energy regimes. Furthermore, our research provided valuable insights into the stability of these defects, both in their neutral and charged states, as well as their corresponding spin characteristics. The findings regarding these two key parameters indicate the potential suitability of native defects in cBN for deployment as quantum bits in quantum technologies.

\begin{acknowledgments}
All the calculations in this work are done at the Phenikaa University's HPC Systems.
The research is funded by Vietnam National Foundation for Science and Technology Development (NAFOSTED) under grant number 103.02-2021.95.
\end{acknowledgments}

\newpage 

\bibliography{cBN_defects}

\end{document}